# Arrest of flow and emergence of activated processes at the glass transition of a suspension of particles with hard sphere-like interactions

W. van Megen, V. A. Martinez and G. Bryant

*Department of Applied Physics, Royal Melbourne Institute of Technology, Melbourne, Victoria 3000, Australia*

Abstract.

By combining aspects of the coherent and self intermediate scattering functions, measured by dynamic light scattering on a suspension of hard sphere-like particles, we show that the arrest of particle number density fluctuations spreads from the position of the main structure factor peak. Taking the velocity auto-correlation function into account we propose that as density fluctuations are arrested the system's ability to respond to diffusing momentum currents is impaired and, accordingly, the viscosity increases. From the stretching of the coherent intermediate scattering function of the colloidal glass we read a quantitative manifestation of the undissipated thermal energy—the source of those, ergodicity restoring, processes that short-circuit the sharp transition to a perfect glass.

In this Letter we report dynamic light scattering measurements that simultaneously expose the arrest of flow and the emergence of activated processes at the glass transition of a suspension of particles with hard sphere-like interactions. Despite numerous studies over about the last two decades [1] the process of vitrification of this and other systems with simple interactions remains poorly understood [2]. The question that continues to be posed is, "What are the mechanisms that cause structural relaxation to slow and resistance to flow to increase so sharply when a liquid is (under-) cooled below its normal freezing temperature?". Interestingly, efforts to answer this question generally ignore one important aspect of the phenomenology – the fact that vitrification is a consequence of the suppression of crystallization. Indeed, one might ask, "What precisely is it about the kinetic path that nucleates the crystal phase, that can be frustrated to such an extent that the solid that emerges is amorphous rather than crystalline?"

To address these issues we combine aspects of the coherent and incoherent (or self) intermediate scattering functions (ISFs), obtained by dynamic light scattering (DLS). Although both ISFs of hard-sphere suspensions have been measured previously [3-5], only recently, as a result of improvements in experimental equipment and procedures, has the data achieved the accuracy required for an analysis that allows expression of the particle number density fluctuations in terms of the single particle displacement. This, in turn, exposes a coupling between viscosity and structural relaxation, over and above that originally proposed by Gesti [6].

We concentrate on volume fractions, $\phi$, in the range approximately between the observed first order freezing transition, at $\phi_f$ (=0.494), and the glass transition (GT), at $\phi_g$ ($\approx$0.565) [7]. It has been established that a small spread in the particle radii (polydispersity) delays nucleation of the crystalline phase long enough to measure the dynamical properties of the under-cooled suspension [8,9]. Irrespective of the delay of nucleation, identification of the phase boundary is essential to distinguish the thermodynamically stable from undercooled (disordered) suspensions.

The particles comprise poly-(methylmethacryate and tri-fluoroethylacrylate) with a stabilizing coating of poly-12-hydroxystearic acid approximately 10nm thick which provides the hard sphere-like interaction. Suspending these in cis-decalin provides samples that, at room temperature, are not only weakly turbid visually but from which multiple scattering of the laser light is negligible [10]. We ascertain the impost of multiple scattering by comparing the intensity correlation functions obtained in auto and cross correlation modes [11]. This suffices for measurement of the coherent ISF. Including some 2% (by volume) of silica particles having the same stabilizing coating and average hydrodynamic radius (R=200nm) as the polymer particles provides samples for which the scattered light has both coherent and incoherent contributions. Temperature variation allows fine tuning of the relative scattering amplitudes of the two species to the point where coherent scattering, ie, scattering from the particle number density fluctuations, is optically suppressed and only the incoherent ISF is measured [4]. Further details of the sample preparation and light scattering procedures are published [4,10].

We begin with the Gaussian limit of the incoherent ISF,

$$F_s(q,\tau) = \exp\left[-q^2 \langle \Delta r^2(\tau) \rangle / 6 \right], \qquad (1)$$

where q is the magnitude of the scattering vector and $\langle \Delta r^2(\tau) \rangle$ the particles' mean-squared displacement (MSD) for delay time $\tau$. The coherent ISF can be expressed in an analogous form as follows,

$$F(q,\tau) = \exp\left[-q^2 w(q,\tau) \right]. \qquad (2)$$

For illustrative purposes we show, in Fig. 1, $\langle \Delta r^2(\tau) \rangle$ and the "width" function, $w(q,\tau)$, for the volume fraction, $\phi=0.53$, about midway between the first order and glass transition values. Diffusive short and long time limits, defined by time windows where $\langle \Delta r^2(\tau) \rangle$ and $w(q,\tau)$ grow in proportion to $\tau$, are commonly considered [12].

We expose non-Fickian, many-body aspects of the density fluctuations by focussing on those points, at delay times $\tau_s$ and $\tau_c(q)$, where $\langle \Delta r^2(\tau) \rangle$ and $w(q,\tau)$, respectively, exhibit their maximum deviations from Fickian behaviour. The deviations are given by the stretching, or non-Fickian, indices,

$$c_s = 1 - \min\left[\frac{d \log \langle \Delta r^2(\tau) \rangle}{d \log \tau}\right] \quad \text{and} \quad c_c(q) = 1 - \min\left[\frac{d \log w(q,\tau)}{d \log \tau}\right]. \qquad (3)$$

They express, on a scale of zero to one, the effect of caging. Alternatively, we read in $c_s$ [or $c_c(q)$], the correlation of forward and backward displacements, incurred by caging, in delay time $\tau_s$ [or $\tau_c(q)$]. So, for a suspension in the limit of infinite dilution, where all particles engage in random Markovian excursions, $c_s = c_c(q) = 0$, and the delay times $\tau_s$ and $\tau_c(q)$ are undefined. For the perfect glass, where all particles are caged, $c_s = c_c(q) = 1$ and $\tau_s$ and $\tau_c(q)$ are infinite. Outside these idealizations $\tau_s$ and $\tau_c(q)$ are the delay times that must be exceeded for tagged particle motion and number density fluctuations of spatial frequency q, respectively, to forget the effects of caging. So, increasing $\tau$ beyond $\tau_s$ and $\tau_c(q)$ results in approach to the (long time) diffusive limit (as illustrated in Fig. 1).

Figs. 2 and 3 show $c_c(q)$ and $\tau_c(q)$, respectively. The variation with q of both $c_c(q)$ and $\tau_c(q)$ decreases with increasing $\phi$. Though $c_c(q)$ retains some q-dependence, $\tau_c(q)$ shows no systematic variation with q at $\phi_g$. Of course, given the stretching of $w(q,\tau)$ at the elevated volume fractions considered here, there is more uncertainty in $\tau_c(q)$ than in $c_c(q)$. Nonetheless, for a given q one generally sees that both $\tau_c(q)$ and $c_c(q)$ increase with volume fraction: as expected, density fluctuations becomes slower and caging more pronounced as $\phi$ increases. Where they can be discerned, minima in $c_c(q)$ and $\tau_c(q)$ indicate that non-Fickian processes are least subject to caging around $q_m$. Moreover, these processes are fastest around $q_m$, in contrast to the short-time diffusive dynamics which are slowest at $q_m$ [12].

Since we have at our disposal previous measurements of the self ISF [4,5] comparisons can be performed of single particle dynamics and particle number density fluctuations. We compare: (i) $c_c(q_m)$ with $c_s$, the value of the non-Fickian index of the MSD (inset Fig. 2), and (ii) $\tau_c(q_m)$ with $\tau_s$, the delay time where the MSD shows its greatest stretching (inset, Fig. 3). Note that $c_s$ converges to $c_c(q_m)$ as $\phi$ approaches $\phi_g$ and, within experimental error, $\tau_s$ coincides with the minimum value, $\tau_c(q_m)$, of $\tau_c(q)$.

From the data in Refs. 4 and 5 we obtain also the MSD for each volume fraction that matches as closely as possible that of a measurement of the coherent ISF. Then for each value of $\tau_c(q)$, where the coherent ISF has its maximum stretching, we read the MSD, $\langle \Delta r^2(\tau_c(q)) \rangle$, at that delay time. This quantity, shown in Fig. 4, represents the mean-squared distance particles must traverse in order for number density fluctuations of wavevector q to lose memory of the effects of caging. [To avoid the possibility of confusion in this unconventional analysis, we emphasise that the q-dependence of the MSD, $\langle \Delta r^2(\tau_c(q)) \rangle$, enters implicitly through the q-dependence of the delay time $\tau_c(q)$.] That these MSDs, $\langle \Delta r^2(\tau_c(q_m)) \rangle \approx 0.15$, are smallest at $q_m$ is consistent with the q dependence of $\tau_c(q)$ seen in Fig. 3. As also anticipated from the behaviour of $\tau_c(q)$, but striking nonetheless, $\langle \Delta r^2(\tau_c(q)) \rangle$ loses its dependence on q with increasing $\phi$.

As shown in previous work [13] and shown in Fig. 4 (inset), the value of the MSD at the delay time, $\tau_s$, of its maximum stretching, shows essentially no systematic variation over the range of volume fractions considered. Since $\tau_c(q_m) \approx \tau_s$ (Fig. 3, inset), $\langle \Delta r^2(\tau_c(q_m)) \rangle$ ($\approx 0.15$) displays the same lack of sensitivity to the volume fraction. This saturation of the MSD hints at an interruption in the decay of density fluctuations of spatial frequency $q_m$ due, presumably, to some structural impediment. As far as we can gauge, from the stretching of the ISFs [Eq. (1)], this interruption, or pause, sets in around the freezing volume fraction, $\phi_f$ [see Fig. 4 of Ref. 14]. Then as $\phi$ increases it is seen to spread to other wavevectors from $q_m$ in both directions. In particular, large scale displacements, generally exposed at lower q, decrease most as $\phi$ approaches $\phi_g$. It appears that at $\phi_g$ number density fluctuations, at least within the measured range of spatial frequencies, $1<qR<5$, all pause at the same MSD, $\langle \Delta r^2(\tau_c(q)) \rangle \approx 0.15$.

For those spatial frequencies for which structural relaxation is interrupted there will be a corresponding interruption to the system's ability to respond to diffusing momentum currents, ie, viscous flow. The local violation of momentum conservation, implied here, is also evident from an absence of or incompatibility with the classical $\tau^{-3/2}$ hydrodynamic "tail" in the velocity auto-correlation function, found for both colloidal [13,14] and Newtonian [15] hard-sphere systems. These studies show, more definitively, that this dynamical anomaly sets in at $\phi_f$. They also show that the structural impediment, whose morphology has been associated with the dynamical heterogeneity [14,15], is a permanent feature of the under-cooled hard sphere system. In other words, dynamical heterogeneity—clusters, within which particles are immobilised, immersed in relatively more mobile material—exposed in microscopic experiments [16] and often used to characterize collective processes in under-cooled fluids [17,18], should be considered perhaps in the context of the first order transition.

While, for a given $\phi$ ($>\phi_f$), the diffusing momentum is blocked by spatial modes in the range, $q_m \pm \delta q$, say, it can still "leak" through those spatial modes outside this range. So the final decay ($\tau > \tau_c(q)$) of density fluctuations in the said range of spatial modes, that no longer couple directly to the diffusing momentum, is delayed by the decay of those that do. So, conversely to the mechanism, proposed by Gesti [6], by which the suspension's (relative) viscosity controls structural relaxation, the impediment to structural relaxation exposed here controls the viscosity. The coupling between flow and structural relaxation is maintained, whatever the mechanism. Accordingly, one anticipates the divergence in the delay times $\tau_c(q_m)$ and $\tau_s$, apparent from Fig. 2 (inset) as $\phi$ approaches $\phi_g$, to be accompanied by a corresponding divergence in the viscosity.

The reasoning above leads to a sharp ergodic to non-ergodic transition to a "perfect" glass, as predicted by the idealised mode coupling theory (MCT) [19]. This theory's predictions describe quantitatively ISFs measured for these suspensions in many significant respects [3,20]. However, the fact that colloidal glasses age [21] is one indication that there is a mechanism we have not considered—the thermal energy left undissipated by the above structural impediment to momentum diffusion. As proposed above, this emerges at $\phi_f$. Then as $\phi$ is increased, and the spread of spatial modes closed to viscous flow increases, one reaches a cross-over where stationary processes have become so slow that the (final) decay of the ISF, in the experimental time window, is effected through non-stationary rather than stationary processes. In these suspensions this cross-over occurs fairly consistently at, or over a narrow range around, $\phi_g \approx 0.57$.

Moreover, were the colloidal glass perfect one would expect $c_c(q)$ and $c_s$ to equal one. One sees from the top panel of Fig. 2 (shown on an expanded scale), that this is not the case. Presumably then, the differences of $c_c(q)$ from one are indicative of those, ergodicity restoring, processes omitted by the idealized MCT theory. Accordingly, we read in the differences of $c_c(q)$ from one a manifestation of the thermal energy left undissipated by the structural impediment to momentum diffusion. Now the q-dependence of $c_c(q)$ indicates that the efficacy of the ergodicity restoring processes is strongest around $q_m$; The system attempts to restore ergodicity through those (local) re-arrangements that must ultimately lead to structures that support lattice modes.

To see how the development of the requisite lattice planes can be frustrated so easily by a modest polydispersity (approximately 6%) we consider results from light scattering studies of crystallization of these suspensions [9]. These reveal, more directly than the DLS experiments, the presence of amorphous clusters of particles. They also show that the clusters are embryonic to crystal nuclei. The clusters' average linear dimension, of some 15 particle diameters, shows little variation with either $\phi$ or polydispersity [9], even though both aspects have a very strong bearing on the time that elapses between the quench and the appearance of Bragg reflections. Of particular significance is the clusters' compactness; For example, the volume fraction within them is approximately 0.61 at the suspension's GT ($\phi \approx 0.565$). So, even a small spread in particle radii, say a few percent of the mean, would incur, by the requirement of exchange of species, significant kinetic impediment to the formation of lattice planes [22]. Accordingly, this local compactness can only impede the local rearrangements that the unrelaxed momenta seek to effect. This addresses the second question posed in the opening paragraph of this Letter.

In conclusion, we offer the following response to the first question; As the suspension's volume fraction is increased from $\phi_f$, partial arrest of number density fluctuations spreads from the position of the main structure factor peak to other wavevectors. Concomitantly, the viscosity increases not just because density fluctuations become slower, as for a system in thermodynamic equilibrium, but also because of a decrease in the number of spatial modes by which thermal energy can dissipate. The thermal energy left undissipated drives activated processes that, inexorably and irreversibly, lead to separation of the (equilibrium) crystal phase. Neglecting the undissipated thermal energy would, in theory, allow density fluctuations of **all** wavevectors to be arrested and the viscosity to diverge.

REFERENCES

1. P.G. Debenedetti, *Metastable Liquids, Concepts and Principles* (Princeton University Press, Princeton, 1996); E. Dhont, *The Glass Transition* (Springer, Berlin, 2001)

2. K.L. Ngai, J. Non-Cryst. Solids **353**, 709 (2007).

3. W. van Megen and S.M. Underwood, Phys. Rev. E, **49**, 4206-4220 (1994).

4. W. van Megen, T.C. Mortensen, S.R. Williams and J. Müller, Phys. Rev. E, **58**, 6073-6085 (1998).

5. W. van Megen, T.C. Mortensen and G. Bryant, Phys. Rev. E **72** 031402 (2005).

6. T. Gesti, J. Phys. C: Solid State Phys. **16**, 5805 (1983).

7. P.N. Pusey and W. van Megen, Nature (London) **320**, 340 (1986).

8. S.I Henderson and W. van Megen, Phys. Rev. Lett. **80**, 877 (1998).

9. H. J. Schöpe, G. Bryant and W. van Megen, J. Chem. Phys. **127**, 084505 (2007); H. J. Schöpe, G. Bryant and W. van Megen, Phys. Rev. Lett. **96**: 175701 (2006).

10. S.M. Underwood and W. van Megen, Colloid Polym. Sci., **274**, 1072-1080 (1996).

11. P.N. Segrè, et al., J. Mod. Optics, **42**, 1929-1952 (1995).

12. P.N. Pusey, in *Liquids, Freezing and the Glass Transition* edited by J.-P. Hansen, D. Levesque and J. Zinn-Justin (North-Holland, Amsterdam, 1991), p. 763.

13. W. van Megen and G. Bryant, Phys. Rev. E **76**, 021402 (2007).

14. W. van Megen, Phys. Rev. E **73**, 020503(R) (2006).

15. S. R. Williams, G. Bryant, I. K. Snook and W. van Megen, Phys. Rev. Lett. **96**, 087801 (2006).

16. A. Kasper, E. Bartsch and H. Sillescu, Langmuir **14**, 5004 (1998); W. Kegel and A. van Blaaderen, Science **287**, 290 (2000); A.S. Keys, et al., Nature (Physics) **3**, 260 (2007).

17. M.D. Ediger, Annu. Rev. Phys. Chem. 51, 99 (2000).

18. D. Chandler, et al., Phys. Rev. E **74**, 051501 (2006); L. Berthier, et al., J. Chem. Phys. **126**, 184504 (2007); G. Szamel, Phys. Rev. Lett. **101**, 205701 (2008).

19. W. Götze, J. Phys.: Condens. Matter **11**, A1 (1999).

20. W. van Megen, Phys. Rev. E **76**, 061401 (2007).

21. V. A. Martinez, G. Bryant and W. van Megen, Phys. Rev. Lett. **101**, 135702 (2008).

22. S.R. Williams, I.K. Snook and W. van Megen, Phys. Rev. E **64**, 021506 (2001); S.R. Williams, C. P. Royall and G. Bryant, Phys. Rev. Lett. 100, 225502 (2008).

LEGENDS

Fig. 1. MSD, $\langle \Delta r^2(\tau) \rangle$, (filled squares), and the width function, $w(q_m,\tau)$ (open squares), versus delay time, $\tau$. Arrows indicate times, $\tau_c(q_m)$ (top) and $\tau_s$. In this and figures below, distances are expressed in units of the particle radius and times in units of the Brownian time ($=R^2/6D_0$). Dashed lines of unit slope indicate diffusive limits.

Fig. 2. (Color on-line) Stretching indices (defined in Eq. 3) versus wavevector for volume fractions indicated. Scale of top panel is 5 times that of bottom panel. The inset shows $c_c(q_m)$ (open diamonds) and $c_s$ (filled diamonds) for all experimental volume fractions.

Fig. 3. (Color on-line) Characteristic delay times for volume fractions indicated in Fig. 2. For clarity not volume fractions are included in main panel. Inset shows $\tau_c(q_m)$ (open diamonds) and $\tau_s$ (filled symbols) for all experimental volume fractions.

Fig. 4. (Color on-line) MSDs at delay times $\tau_c(q)$ versus wavevector for volume fractions indicated in Fig. 2. For clarity not all volume fractions are included in main panel. Inset shows $\tau_c(q)$ at $qR=1.5$ (squares), $qR=3.5$ ($\approx q_m R$) (circles), $qR=4.5$ (triangles), and $\tau_s$ (filled diamonds) for all experimental volume fractions.

Fig. 1

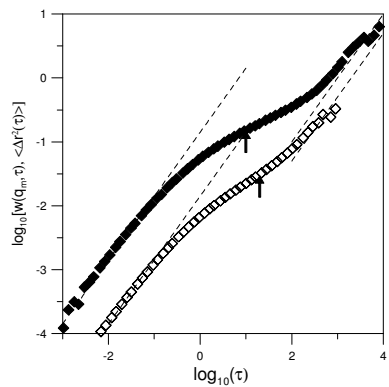

Fig. 2

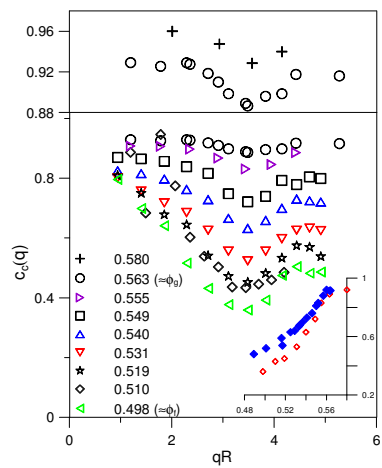

Fig. 3

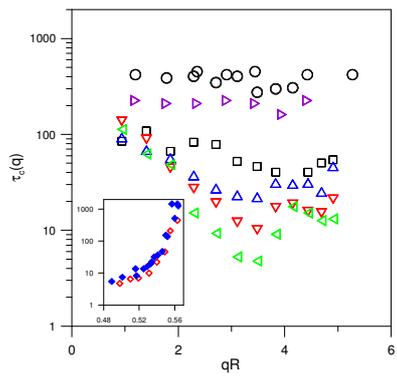

Fig. 4

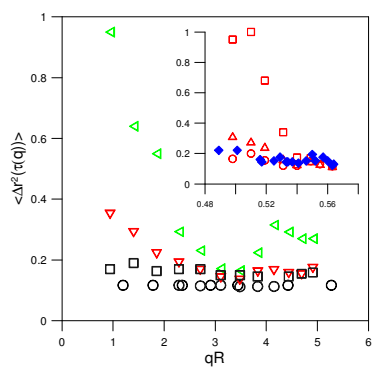